\documentclass[twocolumn,prb,showpacs,preprintnumbers,amsmath,amssymb,floatfix]{revtex4}
\usepackage{graphicx}
\usepackage{amsmath}
\usepackage{amssymb}
\usepackage{color}

\begin{document}

\title{Energy levels of a two-dimensional hydrogen atom with spin-orbit Rashba interaction}

\author{C. Grimaldi}

\affiliation{Max-Planck-Institut f\"ur Physik komplexer Systeme,
N\"othnitzer Srt.38, D-01187 Dresden Germany \\
LPM, Ecole Polytechnique F\'ed\'erale de
Lausanne, Station 17, CH-1015 Lausanne, Switzerland}

%\date{15-09-04}

%\widetext

\begin{abstract}
Electronic bound states around charged impurities in two-dimensional systems with
structural inversion asymmetry can be described in terms of a two-dimensional hydrogen
atom in the presence of a Rashba spin-orbit interaction. Here, the energy levels of
the bound electron are evaluated numerically as a function of the spin-orbit interaction,
and analytic expressions for the weak and strong spin-orbit coupling limits are compared with
the numerical results. It is found that, besides the level splitting due to the lack of
inversion symmetry, the energy levels are lowered for sufficiently strong spin-orbit coupling,
indicating that the electron gets more tightly bound to the ion as the spin-orbit interaction
increases. Similarities and differences with respect to the two-dimensional Fr\"ohlich
polaron with Rashba coupling are discussed.
\end{abstract}
\maketitle

The two-dimensional (2D) hydrogen atom, {\it i. e.}, an electron constrained to move in a
plane and subjected to an attractive Coulomb potential,\cite{kohn,zaslow,yang,dittrich,parfitt}
is a theoretical construction which,
besides being of interest in itself, has also important physical realizations. It can describe indeed
the effect of a charged impurity in 2D systems such as quantum wells and surface states,
or in extremely anisotropic three-dimensional crystals,\cite{kohn} as well as
excitons in semiconductor 2D heterostructures.\cite{parfitt}

The spin-orbit (SO) interaction, arising from the structural and/or bulk inversion asymmetries,
characterizes several of the above mentioned low-dimensional systems,\cite{dassarma} and gives rise to energy
level splittings ranging from a few to hundreds of meV, depending on the material characteristics
(see for example Ref.[\onlinecite{grima2}]).
Furthermore, the possibility of tuning the SO interaction in semiconductor quantum wells by means
of external applied voltages represents the key feature for application in spintronics.
Given this situation, it becomes therefore natural to assess how the properties of a 2D
hydrogen atom are affected by the SO interaction.

Several studies have already been devoted to the effect
of the SO coupling in electrons interacting with central potentials, such as those describing
hard-wall or parabolic quantum dots.\cite{bulgakov,tsitsi,kuan,chakra,chaplik}
However, despite its potential interest for SO coupled low-dimensional systems,
the specific 2D Coulomb problem appears to have been only marginally considered in the literature.\cite{chaplik}
In this Brief Report, the 2D Coulomb problem is numerically solved for an electron interacting with a
Rashba potential, that is the SO coupling arising from structural inversion asymmetry in the
direction perpendicular to the 2D plane.\cite{rashba} It is found that the Rashba interaction removes partially
the initial degeneracy of the 2D hydrogen atom, and the resulting energy levels are two-fold degenerate
due to the time-reversal invariance of the model. Furthermore, it is shown that the SO interaction
renders the electron more tightly bound to the ion, confirming a general trend observed for
other central potentials and for 2D electrons coupled to phonons.

The Hamiltonian for a 2D hydrogen atom in the presence of a Rashba SO potential is as follows
($\hbar=1$):
\begin{equation}
\label{ham1}
H=\frac{\hat{p}^2}{2m_e}-\frac{Ze^2}{r}+\gamma(\hat{p}_x\sigma_y-\hat{p}_y\sigma_x),
\end{equation}
where $\hat{p}_q=-i\partial/\partial q$ is the electron momentum operator ($q=x,y$),
$\hat{p}^2=\hat{p_x}^2+\hat{p_y}^2$, $m_e$ is the electron mass, and $\sigma_x$ and $\sigma_y$ are Pauli
matrices. The last term of Eq.\eqref{ham1} describes the Rashba SO interaction with
coupling parameter $\gamma$. For $\gamma\neq 0$ but zero Coulomb interaction ($Z=0$),
Eq.\eqref{ham1} is easily diagonalized
in the momentum space, and the resulting energy dispersion of the free electron is composed
of two branches $E_{k,\pm}=(k\pm k_R)^2/2m_e-E_R$, where $k_R=m_e\gamma$ is the Rashba momentum.
In the ground state, the electron has energy $E_{k_R,-}=-E_R$, where $E_R=k_R^2/2m_e=m_e\gamma^2/2$.

In the presence of the Coulomb interaction, it is convenient to rewrite Eq.\eqref{ham1}
in polar coordinates:
\begin{equation}
\label{ham2}
H=\left[
\begin{array}{cc}
H_0-\frac{Ze^2}{r} & -\gamma e^{-i\phi}\left(\frac{\partial}{\partial r}
-\frac{i}{r}\frac{\partial}{\partial\phi}\right) \\
 \gamma e^{i\phi}\left(\frac{\partial}{\partial r}+
\frac{i}{r}\frac{\partial}{\partial\phi}\right) & H_0-\frac{Ze^2}{r}
\end{array}\right],
\end{equation}
where
\begin{equation}
\label{h0}
H_0=-\frac{1}{2m_e}\left(\frac{\partial^2}{\partial r^2}
+\frac{1}{r}\frac{\partial}{\partial r}
+\frac{1}{r^2}\frac{\partial^2}{\partial\phi^2}\right)
\end{equation}
is the free electron Hamiltonian. Equation \eqref{ham2} commutes with
the $z$-projection of the total angular momentum $\hat{J}_z=\hat{L}_z+\sigma_z/2$,
where $\hat{L}_z=-i\partial/\partial\phi$, so that the eigenfunctions of \eqref{ham2}
can be chosen to be simultaneously eigenfunctions of $\hat{J}_z$. Since $H$ in polar
coordinates allows for separation of variables, its eigenfunctions have therefore the
following form:\cite{bulgakov,tsitsi,kuan,chakra,chaplik}
\begin{equation}
\label{eigen1}
\Psi_j(r,\phi)=\left[
\begin{array}{c}
f_j^{-}(r)e^{i(j-1/2)\phi} \\
f_j^{+}(r)e^{i(j+1/2)\phi}
\end{array}\right]
\end{equation}
where $j=\pm 1/2,\pm 3/2,\ldots$ are the eigenvalues of $\hat{J}_z$.
The lack of spatial inversion symmetry induced by the
presence of the Rashba interaction lowers the symmetry of $H$ when $\gamma\neq 0$. As
shown below, this will induce a splitting of the energy levels compared to the
case when $\gamma=0$. Note however that $H$
commutes with the time-reversal operator $\hat{K}=i\sigma_y \hat{C}$, where $\hat{C}$ is the
operation of complex conjugation, so that $\Psi_j$ and its Kramer conjugate $\hat{K}\Psi_j$
have the same energy. This implies that, since $\hat{J}_z\Psi_j=j\Psi_j$ and
$\hat{J}_z\hat{K}\Psi_j=-j\hat{K}\Psi_j$, the energy spectrum of $H$ is invariant under
the change $j\rightarrow -j$.

For bound states, the Schr\"odinger equation $H\Psi_j=E\Psi_j$ is rewritten by introducing
$q_0^2=-2m_eE$ and the dimensionless radial variable $\rho=2q_0r$. By using Eq.\eqref{eigen1}
one therefore finds
\begin{align}
\label{a1}
\left[\frac{d^2}{d\rho^2}+\frac{1}{\rho}\frac{d}{d\rho}-
\frac{(j-1/2)^2}{\rho^2}+\frac{\lambda}{2q_0\rho}-\frac{1}{4}\right]&f_j^{-} \nonumber \\
+\frac{k_R}{q_0}\left(\frac{d}{d\rho}+\frac{j+1/2}{\rho}\right)&f_j^{+}=0, \\
\label{b1}
\left[\frac{d^2}{d\rho^2}+\frac{1}{\rho}\frac{d}{d\rho}-
\frac{(j+1/2)^2}{\rho^2}+\frac{\lambda}{2q_0\rho}-\frac{1}{4}\right]&f_j^{+} \nonumber \\
-\frac{k_R}{q_0}\left(\frac{d}{d\rho}-\frac{j-1/2}{\rho}\right)&f_j^{-}=0,
\end{align}
where $\lambda=2m_eZe^2$.
The first terms in Eqs.\eqref{a1} and \eqref{b1} represent the differential equations for the radial
wave function of the usual 2D Coulomb problem ({\it i. e.}, without SO coupling) with
quantum numbers $j-1/2$ and $j+1/2$, respectively.\cite{zaslow,yang}
Apart from a normalization constant,
their solutions are of the form
$R_{N,j\pm 1/2}(\rho)=\exp(-\rho/2)\rho^{\mid j\pm 1/2\mid}L_N^{2\mid j\pm 1/2\mid}(\rho)$
where $N=0,1,2,\dots$ is the
radial quantum number and $L_N^{2\mid j\pm 1/2\mid}(\rho)$ are Laguerre polynomials.\cite{zaslow,parfitt}
The corresponding energy levels are
\begin{equation}
\label{eigen2}
E_{N,j\pm 1/2}^0=-\frac{\eta/4}{\left(N+\mid j\pm \frac{1}{2}\mid+\frac{1}{2}\right)^2},
\end{equation}
where $\eta=2m_eZ^2e^4=\lambda^2/2m_e$. By introducing the principal quantum number
$n=N+\mid j\pm 1/2\mid=0,1,2\ldots$, with $\mid j\pm 1/2\mid \leq n$, one infers that each level
with fixed $n$ has energy $-\eta/(2n+1)^2$ with degeneracy $2(2n+1)$.

\begin{figure}[t]
\protect
\includegraphics[scale=0.7,clip=true]{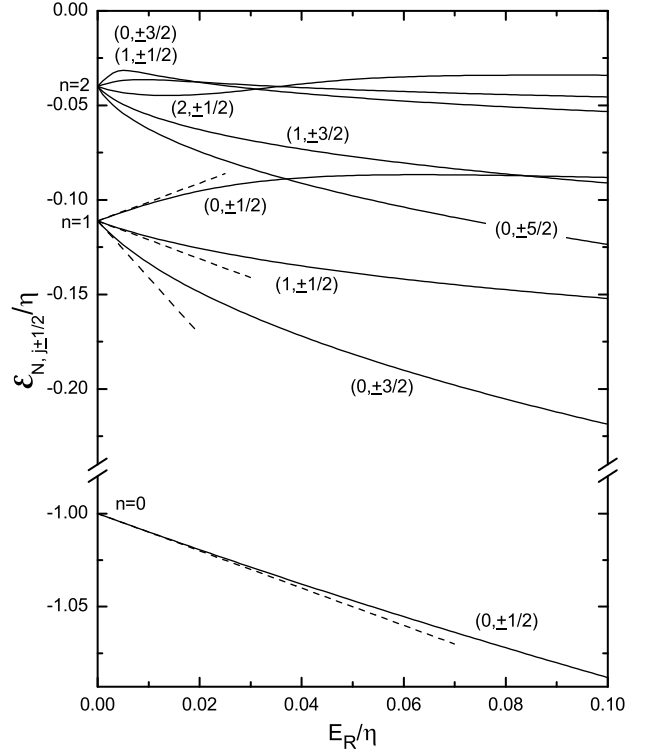}
\caption{Energy levels obtained from a numerical solution of Eqs.\eqref{a3} and \eqref{b3} (solid lines)
as a function of the Rashba energy $E_R$.
The dashed lines are the analytic results of Eq.\eqref{eigen3} for the weak SO limit.
All energy levels are shifted with respect to the ground state of the free electron with SO
interaction [Eq.\eqref{eigen4}]. The different levels are labeled by the principal quantum
number $n$, reported in the left axis, and by the radial quantum number $N$ and the total
angular momentum in the $z$ direction $j$ reported in parentheses.} \label{fig1}
\end{figure}

For nonzero SO coupling, it is natural to expand the
radial functions $f_j^{\pm}$ in terms of $R_{N,j\pm 1/2}(\rho)$. By keeping in
mind that the total wave function $\Psi_j$ must be also eigenfunction of $\hat{J}_z$, one has:
\begin{equation}
\label{a2}
f_j^{\pm}(\rho)=e^{-\rho/2}\rho^{\mid j\pm 1/2\mid}\sum_{N=0}^\infty A_{N,j}^{\pm}L_N^{2\mid j\pm 1/2\mid}(\rho).
\end{equation}
By substituting the above expansions in Eqs.\eqref{a1} and \eqref{b1}, and by making use of the
properties of the Laguerre polynomials,\cite{GR} one arrives at the following iterative system of
equations
\begin{align}
\label{a3}
&\left(\frac{\lambda}{2q_0}-\frac{1}{2}-\mid j-\frac{1}{2}\mid-N\right)A_{N,j}^{-}
+\frac{k_R}{2q_0}\mathcal{C}_{N',N}^j A_{N',j}^{+} =0,\\
\label{b3}
&\left(\frac{\lambda}{2q_0}-\frac{1}{2}-\mid j+\frac{1}{2}\mid-N\right)A_{N,j}^{+}
-\frac{k_R}{2q_0}\mathcal{C}_{N',N}^{-j}A_{N',j}^{-} =0,
\end{align}
where
\begin{align}
\mathcal{C}_{N',N}^j=&\theta(j)[(N+2j+1)(N+2j)\delta_{N',N}\nonumber \\
&-N(N-1)\delta_{N',N-2}]\nonumber \\
&+\theta(-j)[\delta_{N',N+2}-\delta_{N',N}].
\end{align}
The values of $q_0$, and so the energy levels $E=-q_0^2/2m_e$, satisfying Eqs.\eqref{a3} and
\eqref{b3} can be easily obtained analytically in the limit of weak SO coupling. It suffices to
recognize that decoupling Eqs.\eqref{a3} and \eqref{b3} leads to two iterative equations of the form
$a_\pm A_{N,j}^{\pm}+b_\pm A_{N-2,j}^{\pm}+c_\pm A_{N+2,j}^{\pm}=0$, whose solutions in the weak
SO limit are determined simply by the condition $a_\pm=0$, since $b_\pm$ and $c_\pm$ are both of
order $(k_R/\lambda)^2$. Up to order  $(k_R/\lambda)^2$ the coefficients $a_\pm$ are given by
\begin{align}
\label{apm}
a_\pm=&\frac{\lambda}{2q_0}-\left(N+\mid j\pm \frac{1}{2}\mid+\frac{1}{2}\right)\nonumber \\
&\mp\left(\frac{2k_R}{\lambda}\right)^2\left(N+\mid j\pm \frac{1}{2}\mid+\frac{1}{2}\right)^3
\left(j\mp \frac{1}{2}\right),
\end{align}
so that the energy levels $E_{N,j\pm 1/2}$ of the weak SO interacting case are
\begin{equation}
\label{eigen3}
E_{N,j\pm 1/2}=E_{N,j\pm 1/2}^0\pm 2 j E_R-E_R,
\end{equation}
where $E_{N,j\pm 1/2}^0$ is the energy spectrum for zero SO interaction given in Eq.\eqref{eigen2}.
From Eq.\eqref{eigen3} one sees therefore that the $2(2n+1)$-fold degeneracy for zero SO
coupling is lifted when $\gamma\neq 0$ and that each level is splitted into $2n+1$ levels, each
two-fold degenerate. The remaining degeneracy is due to the time-reversal invariance of $H$
and can be removed by adding a time-reversal breaking term in the Hamiltonian such as a magnetic field.
Note also that Eq.\eqref{eigen3} could be obtained by making use of the method described in
Ref.[\onlinecite{aleiner}].

A comparison between the weak SO coupling result \eqref{eigen3} (dashed lines) and
the energy levels obtained by a numerical solution of Eqs.\eqref{a3} and \eqref{b3} (solid lines)
is reported in Fig.\ref{fig1} for a few energy levels. The principal quantum number values $n$
are reported in the left axis, while the radial and total angular momentum
quantum numbers $N$ and $j$ are indicated in parentheses.
What is plotted in Fig.\ref{fig1} is actually the
quantity
\begin{equation}
\label{eigen4}
\mathcal{E}_{N,j\pm 1/2}=E_{N,j\pm 1/2}+E_R,
\end{equation}
that is the energy spectrum shifted with respect to the ground state $-E_R$ of the free electron
coupled to the SO potential.
As it is shown in the figure, the ground state (identified by quantum
numbers $n=0$, $N=0$, and $j=\pm 1/2$) has its energy lowered by the SO interaction, demonstrating that
the electron gets more tightly bounded as $E_R$ increase. This holds true also for the higher energy
levels which, besides being splitted by the Rashba interaction, have their energies
lowered for sufficiently large $E_R$ values, as it apparent for most of the levels
plotted in Fig.\ref{fig1}. For states like $n=1$ $(0,\pm 1/2)$ and $n=2$ $(2,\pm 1/2)$ one needs
$E_R/\eta \gtrsim 2$ before reaching net energy levels lower than the zero SO limit.
This is of course unattainable since, for an unscreened charged impurity, $\eta$ is of the order of
one Ry, while the maximum value of $E_R$ to date is of about $0.2$ eV.\cite{ast} The relevant values
of $E_r/\eta$ are therefore lower than about $0.01$ for which, however, the perturbative
result \eqref{eigen4} is in quantitative good agreement with the numerical solution
plotted in Fig.\ref{fig1}.

\begin{figure}[t]
\protect
\includegraphics[scale=0.7,clip=true]{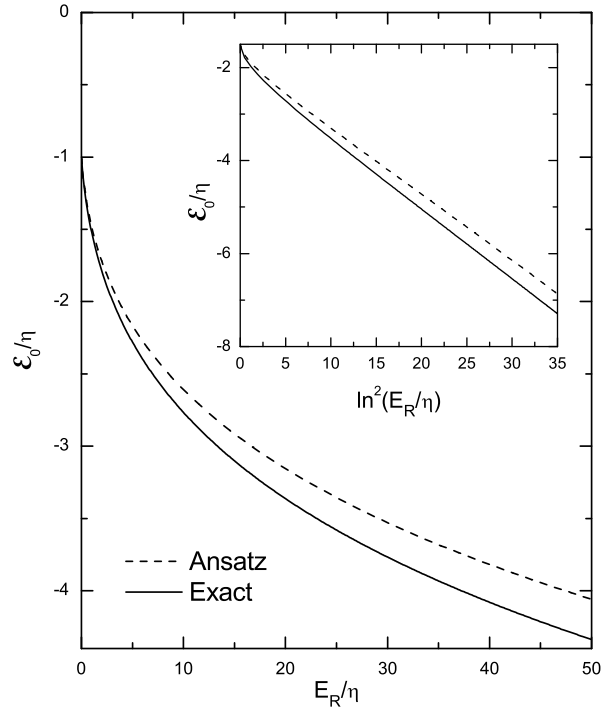}
\caption{Comparison between the exact ground state energy obtained from a numerical
solution of Eqs.\eqref{a3} and \eqref{b3} (solid line) and the variational calculation
with the ansatz wave function \eqref{ansatz1} (dashed line). Inset, the same results
plotted as a function of $\ln^2(E_R/\eta)$.} \label{fig2}
\end{figure}

Let us now discuss the relevance of the above results with respect to
a different but somewhat related problem: that of a 2D electron strongly
coupled to longitudinal optical phonons in the presence of a SO Rashba potential
(Fr\"ohlich-Rashba model).\cite{grima} To this end it is useful to compare the
exact numerical solutions of Eqs.\eqref{a3} and \eqref{b3} with a variational calculation of the
ground state energy $\mathcal{E}_0$ obtained from the following electron trial wave function:
\begin{equation}
\label{ansatz1}
\Psi_0(r,\phi)=\mathcal{A}e^{-ar}\left[
\begin{array}{l}
J_0(br) \\
J_1(br)e^{i\phi}
\end{array}\right],
\end{equation}
where $a$ and $b$ are variational parameters, $J_0$ and $J_1$ are Bessel functions, and $\mathcal{A}$
is a normalization constant.
The above form for $\Psi_0(r,\phi)$ was introduced in Ref.[\onlinecite{grima}] in order to find an upper bound
for the ground state of a the Fr\"ohlich-Rashba model. By using Eq.\eqref{ansatz1}, the energy functional
$\mathcal{F}=\langle\Psi_0\mid (H+E_R)\mid\Psi_0\rangle$ can be found analytically
\begin{equation}
\label{ansatz2}
\mathcal{F}=\frac{a^2}{2m_e}+\frac{(b-k_R)^2}{2m_e}
+\frac{\lambda a}{m_e}\!\left[1+\frac{K(ib/a)(b/a)^2}{K(ib/a)-E(ib/a)}\right],
\end{equation}
where $K$ and $E$ are complete elliptic integrals of the first and second kind, respectively.
Numerical minimization of Eq.\eqref{ansatz2} with respect to $a$ and $b$ provides an upper bound
$\mathcal{E}_0$ for the ground state energy. This is plotted in Fig.\ref{fig2} by the dashed line and
compared with the exact ground state energy (solid line) obtained from Eqs.\eqref{a3} and \eqref{b3}.
It is seen that the simple ansatz \eqref{ansatz1} reproduces fairly well the lowering of the ground
state energy as $E_R$ increases. In particular, by expanding Eq.\eqref{ansatz2} for small values
of $E_R$ compared to $\eta=\lambda^2/2m_e$ it turns out that $\mathcal{E}_0\simeq -\eta-E_R$, which coincides with
Eqs.\eqref{eigen3} and \eqref{eigen4} for $n=0$, $N=0$ and $j=1/2$. For very large values of $E_R/\eta$,
Eq.\eqref{ansatz2} has the limiting form
\begin{equation}
\label{ansatz3}
\mathcal{F}=\frac{a^2}{2m_e}+\frac{(b-k_R)^2}{2m_e}+\frac{\lambda a}{m_e}\ln \!\left(\frac{ea}{4b}\right),
\end{equation}
where $e$ is the Neper number and should not be confused with the electron charge. Equation \eqref{ansatz3}
is minimized by setting $b=k_R$ and, within logarithmic accuracy, $a\simeq\lambda\ln(4k_R/\lambda e^2)$.
Hence, the corresponding asymptotic upper bound for the ground state energy reduces to:
\begin{equation}
\label{ansatz4}
\mathcal{E}_0\simeq -\eta\ln^2\!\left(\frac{4k_R}{\lambda e^2}\right)
=-\frac{\eta}{4}\ln^2\!\left(\frac{16E_R}{\eta e^4}\right),
\end{equation}
indicating that for $E_R\rightarrow\infty$ the ground state energy gets indefinitely lowered
by following a squared logarithmic dependence on $E_R$.
This result is confirmed in the inset of Fig.\ref{fig2}, where both the exact result (solid line)
and the numerical minimization of \eqref{ansatz2} (dashed line) reduce to straight lines
when plotted as a function of $\ln^2(E_R/\eta)$.

The functional dependence of the ground state energy on the SO
coupling shown in Eq.\eqref{ansatz4} was originally obtained by a different method in
Ref.[\onlinecite{chaplik}] where, however, a cutoff parameter was introduced to prevent a
diverging result. As further noted in Ref.[\onlinecite{chaplik}], a squared logarithmic
behavior characterizes also the ground state energy of the three-dimensional (3D) hydrogen atom in
an extremely strong magnetic field $H$,\cite{landau,kleinert} supporting the interpretation that a 2D
electron in the presence of a strong Rashba SO interaction behaves effectively as a one-dimensional (1D)
particle. The correspondence between 1D-like behavior and strong Rashba interaction has been
notices also for bound states of 2D electrons in short range central
potentials,\cite{chaplik,galstyan} as well as
for 2D electrons coupled to phonons.\cite{grima,grima2}
Such correspondence however does not appear to have universal validity. In fact
when the ansatz \eqref{ansatz1} is used
in the 2D Fr\"ohlich-Rashba model, the asymptotic strongly-coupled polaron ground state energy
for $E_R\rightarrow\infty$ does not decreases indefinitely as Eq.\eqref{ansatz4} but, rather, it
reaches a minimum finite value.\cite{grima} This is in striking contrast with the 3D strongly
coupled Fr\"ohlich polaron in a strong magnetic field, whose ground state energy has a squared
logarithmic functional form as the 3D hydrogen atom for $H\rightarrow\infty$,\cite{polarH}
due to the 1D confining effect of the magnetic field on the electron motion.


\begin{thebibliography}{99}

\bibitem{kohn}
W. Kohn and J. M. Luttinger, Phys. Rev. {\bf 98}, 915 (1955).

\bibitem{zaslow}
B. Zaslow and M. E. Zandler, Am. J. Phys. {\bf 35}, 1118 (1967).

\bibitem{yang}
X. L. Yang, S. H. Guo, F. T. Chan, K. W. Wong, and W. Y. Ching,
Phys. Rev. A {\bf 43}, 1186 (1991).

\bibitem{dittrich}
W. Dittrich, Am. J. Phys. {\bf 67}, 768 (1999).

\bibitem{parfitt}
D. G. W. Parfitt and M. E. Portnoi, J. Math. Phys. {\bf 43}, 4681 (2002).

\bibitem{dassarma}
I. \v{Z}uti\'{c}, J. Fabian, and S. Das Sarma,
Rev. Mod. Phys. {\bf 76}, 323 (2004).

\bibitem{grima2}
E. Cappelluti, C. Grimaldi, and F. Marsiglio,
Phys. Rev. Lett. {\bf 98}, 167002 (2007);
Phys. Rev. B {\bf 76}, 085334 (2007).

\bibitem{bulgakov}
E. N. Bulgakov and A. F. Sadreev, JETP Lett. {\bf 73}, 505 (2001).

\bibitem{tsitsi}
E. Tsitsishvili, G. S. Lozano, and A. O. Gogolin,
Phys. Rev. B {\bf 70}, 115316 (2004).

\bibitem{kuan}
W. H. Kuan, C. S. Tang, and W. Xu, J. Appl. Phys. {\bf 95}, 6368 (2004).

\bibitem{chakra}
P. Pietil\"ainen and T. Chakraborty, Phys. Rev. B {\bf 73}, 155315 (2006).

\bibitem{chaplik}
A. V. Chaplik and L. I. Magarill, Phys. Rev. Lett. {\bf 96}, 126402 (2006).

\bibitem{rashba}
Yu. A. Bychkov and E. I. Rashba, JETP Lett. {\bf 39}, 78 (1984).

\bibitem{GR}
I. Gradshteyn and I. Ryzhik, {\it Table of Integrals, Series, and Products},
(Academic, San Diego, 1994).

\bibitem{aleiner}
I. L. Aleiner and V. I. Fal'ko, Phys. Rev. Lett. {\bf 87}, 256801 (2001).

\bibitem{ast}
C. R. Ast, J. Henk, A. Ernst, L. Moreschini, M. C. Falub, D. Pacil\'{e},
P. Bruno, K. Kern, and M. Grioni, Phys. Rev. Lett. {\bf 98}, 186807 (2007).

\bibitem{grima}
C. Grimaldi, Phys. Rev. B {\bf 77}, 024306 (2008).

\bibitem{landau}
L. D. Landau and E. M. Lifshits, {\it Quantum Mechanics} (Pergamon, Oxford, 1977).

\bibitem{kleinert}
H. Kleinert, {\it Path Integrals in Quantum Physics, Statistics, Polymer Physics,
and Financial Markets} (World Scientific, Singapore, 2006).

\bibitem{galstyan}
A. G. Galstyan and M. E. Raikh, Phys. Rev. B {\bf 58}, 6736 (1998).

\bibitem{polarH}
see for example: F. M. Peeters and J. T. Devreese, Phys. Rev. B {\bf 25}, 7281 (1982);
N. Tokuda and Hatsuhiro, J. Phys. C: Solid State Phys. {\bf 20}, 3021 (1987).



\end{thebibliography}
\end{document}